# A Framework of Value Exchange and Role Playing in Web 2.0 WebSites

**Enas M. Al-Lozi, PhD**
Assistant Professor of Information Systems
Al-Zaytoonah University of Jordan
Department of Management Information Systems

Enas.al-lozi@zuj.edu.jo

**Mutaz M. Al-Debei, PhD**
Assistant Professor of Information Systems
and Computing
The University of Jordan
Department of Management Information Systems
m.aldebei@ju.edu.jo

*ABSTARCT*

*Digitally engaged communities can be described as communities created and evolved within Web 2.0 Websites such as Facebook, Bebo, and Twitter. The growing importance of digitally engaged communities calls for the need to efficiently manage the building blocks of sustaining a healthy community. The initial operation of any digitally-engaged community depends on the existence of its own members, the beneficial values created and exchanged, and the relationships interlinking both. However, the level of contribution and involvement might vary depending on the benefits being gratified from engaging in such communities. In other words, motivations for participating and getting involved are purposive; individuals are driven into joining and /or taking part in any digitally-engaged network for capturing and purtaining certain beneficial values. Accordingly, this paper proposes a framework that classifies the values created and exchanged within these communities as well as the roles adopted and played by users of these communities. Utilizing ethnography as the primay methodological strategy to study Bebo digitally-engaged community, this research identifies five different roles of users: Newbie, Lurker, Novice, Insider, and Leader. Moreover, the research also identifies five value elements that could be captured by different users: Social, Hedonic, Epistemic, Gift, and Utilitarian. The results of this study provides insights for decision and policy makers, service providers, and developers; as it inspires them in knowing and meeting the needs and values of participants based on the roles adopted by users.*

# 1   INTRODUCTION

Along with the wide adoption of Web 2.0 technologies, Digitally Engaged Communities (DECs) are growing exponentially. Therefore, there has been a recent rise of interest in studying such communities. Digitally engaged communities are referred to by terms such as *online communities*, *virtual communities, web communities,* and *social networks*. Such communities can be described as groups of people who conduct open activities of shared practices based on similar interests through Internet and other technologies so as to overcome constraints of time and space (Jin et al., 2007). Nowadays, DECs are the lifeblood of the Internet; the medium that created an online environment for people to get together in a more accessible way. The initial operation of such communities depends on the ongoing participation, and engagement of its own members, as the intended purpose behind them cannot be achieved without the presence of dedicated interactants ensuring an effective functioning of the community. This is because, otherwise, it would simply be a cyberspace of outdated contents rather than an ongoing source of value creation and exchange.

The advent of the third generation Internet-based broadband and other Web-based technologies have transformed the nature of social interacting. Indeed, the rapid growth of network access, and convergence of a faster medium of computer-mediated networking opened opportunities for exchanging value between different parties. This proliferation of low cost access enticed people to manage their social lives online. Therefore, the wide spread of DECs and their mass adoption is traced back to the increasing number of Internet users connected all over the globe. A report by ComScore, for example, indicated that the estimate number of Internet users for the year 2008 counted nearly 1.5 billion Web surfers worldwide, with over 36.7 million within the UK (Schonfeld, 2009).

As the Internet has shifted the boundaries of human interaction, communities have extended to a broader geographical context and more users are now joining DECs. Retrospectively, there is an emerging need to understand interactions at a deeper level (Nolker and Zhou, 2005; Al-Lozi and Papzafeiropoulou, 2010; Al-Lozi, 2011) including th need for a further investigation of values exchanged within DECs, as well as an in-depth illustration of the interrelationships between the roles adopted by users and the values gratified within each role.This is indeed vital so as to improve our understanding of human-human and human-information relationships that can lead to a more effective use of the space. Notwithstanding and despite the fact that there is considerable research on DECs, yet there is still no comprehensive understanding of the types of roles adopted by users along with the values exchanged in such communities and hence; this paper is devoted to examine this issue.

Related literature has embarked upon DECs from multi-disciplined points of view. Within the literature, there have been some attempts to examine types of values exchnaged in DECs, but these efforts have tackled the phenomenon in a more general sense without giving much attention to precisely identifying those values (See for example Chiu et al., 2006; Wang et al., 2008). Furthermore, regarding the roles adopted by users in DECs, and although one can find some research in this area (See for example Preece, 2000; Ridings et al., 2006; Li et al., 2008) yet these roles seem to be underestimated. A coherent classification of them is still absent as researchers have tended to focus mainly on examining a particular role in most cases, or just coming across different roles, rather than investigating them throroughly and cohesively.

Aiming to fulfil this gap, this research is established to investigate ***Why and how would users participate in digitally engaged communities?"***.  This can be further analyzed as: *"What are the values exchanged in digitally engaged communities?", "What are the roles adopted by users of digitally engaged communities so as to fulfil the desired needs and wants?".* The remainder of this paper is organized as follows: Section 2 provides an overview on DECs. Next, section 3 exemplifies the methodology applied. The classifications of values exchanged in DECs are explained in section 4.1. Further on,  users' roles are clarified in section 4.2. Section 5 finally discusses the conceptual framework and the study's concluding remarks and implications.

## 2 WEB 2.0 WEBSITES: DIGITALLY ENGAGED COMMUNITIES

There is no universal definition for DECs. Wilson (2001) argues that as DECs are shifting from a geographical centric to a socialising focus, they are becoming extremely difficult to define. Nonetheless, everyone nowadays seems to have an idea of the hallmarks forming such communities. One of the most widely cited definition of DECs is that of Preece (2000). She argues that a DEC consists of: *People, Purpose, Policies, and the Computer Systems.* She explains that any community is created by a group of people networking together, interacting publicly, sharing similar needs, and governed through implicit set of protocols guiding their interactions. Preece also indicates that this kind of digital relationship needs to be mediated by the support of a technological facilitator. Hence, one can argue that DECs are web-based networks of interpersonal ties connecting people socially, and allowing them to (1) create a sense of belonging and construct an online profile within a bounded system, and (2) articulate a list of other online contacts with whom they establish relationships and connections.

The convergence of a cheaper and faster web-based medium opened opportunities for global networking where exchanging value is the main stream for a healthy community. The massive availability of DECs has deepened the velocity of transactions and fostered interactional density. Due to that, social ties are shifting from linking people in particular places to linking people at any place. However, the initiation of DECs has not been for the sake of their own; as they mostly support the connection of shared interests and views. For example some communities have been emerged with an intention of building relationships (e.g MySpace), enhancing friendships (e.g Facebook), and pertaining emotional/health support (e.g Bebo Bewell), while others have been launched for learning reasons (e.g Pearson), and music sharing purposes (e.g Bebo). On a general level, according to Peck et al.*,* (2007), DECs can be categorized into five main classes:

- **Person Oriented:** This type of communities focus on individuals and their social interactions (e.g. Facebook).
- **Professional:** Professional communities or Communities of Practice (CoP) are communities of knowledge creation and exchange within the boundaries of a specialsed network (e.g. LinkedIn).
- **Media-Oriented:** Communities that focus on the creation, distribution and consumption of user-generated multi-media content, such as videos, music, and photos (e.g.YouTube).
- **Virtual World:** 3D Communities with multimedia tools and applications to enhance user-generated content that is owned by its own members and users (e.g. Second Life).
- **Mobile:** Communities that allow easy access, and make it possible to have direct and indirect contact with the community on the move. Where news, and updates are checked simply through any hand held device installed with web-based applications (e.g.Twitter).

In spite of the purpose behind each community, most share common participative features; i.e. interactants who form impressions through customized personal profiles (Arguello et al., 2006; Dwyer et al., 2007). Such profiles reflect self-presentational behaviours as members share personal information and upload it for contacting purposes, whether via their online profiles in one-to-one basis (much like an email), or in a more public and multi-lateral manner.

## 3 RESEARCH METHODS

The focus of the study is to understand the roles of users in Web 2.0 Websites as well as the nature of value elements they capture. As we are examing the phenomenon of digitally engaged communities from a social and cultral perspectives, a qualitative approach is deemed fitting (See Avison and Pries-Heje, 2005; Denzin and Lincoln, 1994). Amongst the different qualitative approaches, the interpretive research is selected as it is the most appropriate method for the purpose of this study. This is because, as argued by Walsham (1993), interpretive research methods aim at "understanding of the context of the information system and the process whereby the information system influences and is influenced by the context" (p. 4-5). We consider DECs a complex phenomenon and thus an interpretive research is deemed fitting so as to understand the context of such systems and how they influence and influenced by the context. Noticeably, interpretive research is used as an approach that can assist in understanding complex phenomena related to the use of IS in general (Walsham, 1995). In the context of this study, information systems are defined as web-based technologies, their providers (i.e. organizations) and their digitally engaged communities (i.e. people).

The employed methodological strategy in this interpretive research is ethnography given its fit to provide descriptions and a depth of understanding of the human society, community, or culture. of. Hammersley and Atkinson (1993) define ethnography as "a descriptive account of a community or culture". Ethnography can also be described as observational investigation that refers to a field work conducted by investigators who live with and live like those who are studied. Given that this research is examining communities within Web 2.0 Websites from a social and cultral standpoints, ethnography seems to be suitable (Correll, 1995; Hine, 2000). Indeed, using ethnography to examine online communities is common within the IS research (e.g. Nonnecke and Preece, 2000; Schoberth *et al.*, 2003; and DeSouza and Preece, 2004).

The community examined in this research is Bebo (i.e. "Blog Early, Blog Often") social networking platform. Although Bebo has been established just in 2005, the number of its users exceeds 40 million members world-wide (BBC News, 2008). This however makes it one of the largest and fastest growing social networks. It is a digitally engaged community consisting of over (80) groups, sub-groups, and sub sub-groups. Each group serves certain social and other needs including but not limited to mental and other health support, crime prevention, social care, and music and talents share.

Given that this study takes place in a digital community, ethnography here is referred to as "online ethnography" (e.g. Correll, 1995), or "virtual ethnography" (e.g. Hine, 2000). Following online ethnography, the authors were taking more of participative roles rather than observing, where more engagement took place in the virtual space. Authors otherwise referred to as ethnographers, were living among Bebo users over (18) months, and participating in daily activities while maintaining objectivity. In this ethnographic study, the primary source of data is collected through participant observations as it is regarded a core ethnographic technique where researchers participate in observing the behaviour under examination without influencing the patterns of it (Myers, 1999; Sade Beck, 2004). That is direct, first-hand observation of members' daily behaviors including informal conversations and long-term engagement where (1114) messages out of (12) sub-groups were analyzed following content analysis techniques (see Al-Debei and Avison, 2010).

## 4 FINDINGS AND ANALYSIS

**4.1 VALUE CREATION IN DIGITALLY ENGAGED COMMUNITIES**

Digitally engaged communities cannot survive without user involvement and participation in terms of generating content and social interacting. Indeed, DECs need members if they are to be successful. What are most important are values the community offers. These values are created and exchanged by the community's own users. For the past couple of years, millions of people have turned daily to DECs for diverse information-seeking and other communication activities. A great number of users, however, appeared to be passive information consumers (Fichter, 2005; Totty, 2007) without any active involvment. Over time, many of those would assume an additional role and become active content contributors, and thus generating beneficial values (Goodnoe, 2006).

Behind any level of digital participation, are numerous classes of value exchanged amongst users. As DECs offer a wide range of publicly transferred benefits, people join them to fulfil personal needs, whether individually-oriented, or community-oriented. Therefore, participation is purposive and the level of involvement varies depending on the purpose behind joining them. Thus, we postulate that the successful operation of any DEC depends to a large extent on its growing values communicated to its users. The applied analysis in this research reveals that value elements in DECs can be classified as: *Social*, *Hedonic*, *Epistemic*, *Gift*, and *Utilitarian* value elements.

*4.1.1 SOCIAL VALUE*

Social value is one of the most important values captured in DECs (Arguello et al., 2006; Jin et al., 2007). It concerns the utility derived from user's association with certain social groups, and eventually could be broken down into *Emotional*, *Networking*, *Self-Esteem*, and *Self-Discovery* needs. These needs however are maintained through the interpersonal relationships among interactants.

Many individuals join online networks desperately seeking for **Emotional** support in different aspects of life. Calls could be for help and advice in health issues (e.g Bebo/Bewell), mental conflicts (e.g the Samaritans), and decisional support matters (e.g Bebo/Beenriched). Such users might lack the opportunity of getting this support in real life, and somehow been dragged into isolation (Malooney-Krichmar and Preece, 2005). Therefore, they are encouraged into finding an accessible substitute. The anonymity in these communities also increases the calls for community assistance, as social value seekers are offered opportunities to receive emotional support in a climate of trust and empathy (Johnson and Ambrose, 2006).

The value of **Networking** is another goal for many social networking sites. Interactants tend to bond and maintain relationships. It can be described as the process by which members act toward or respond to one another; i.e. social interacting (Chen and Huang, 2007). Networks of socialising can take the form of online that might extent to further boundaries of offline relationships, or enhancing offline relationships via online networks. DECs are all about social interaction where communication is the foundation of such relationships (Ridings and Gefen, 2004).

The value of **Self-Esteem** offered by platforms of DECs enables users to open up and get a feeling of togetherness through interaction with other community members. Participating in groups and events gives members the feeling of existence and being connected. The creation of groups and the contribution to discussions can help establish a certain reputation, which according to the theory of human motivation (Maslow, 1943) represents the outer self-esteem need and thereby enable them to feel internally important (e.g. inner self-esteem). Generally speaking, the esteem needs both on the outer and inner levels to be satisfied from DECs are (1) the need of respect of others, the need for status, recognition, attention, appreciation, even dominance, and (2) the need for self-respect, including feelings such as confidence, achievement, independence, and freedom (Janzik and Herstatt, 2008).

As the last recognizable social value, **Self-Discovery** is defined as: "a sense of emotional involvement with the group" (Bagozzi and Dholakia 2002, p.11), or as described by Blanchard and Markus (2002) it signifies a "sense of community". Joining a group creates a sense of attachment to that group; as long as one's certain needs are satisfied (Dholakia et al., 2004). In DECs, despite the lack of face-to-face interaction, human feelings including attachment, obligation, relationship, identity, and support are important dimensions captured in the sense of belonging to a community (Blanchard and Markus, 2004). Thus, we beleive that the stronger the sense of community belonging individuals conquers, the more they are likely to take more of an active role.

*4.1.2 HEDONIC VALUE*

Hedonic values highlight three personal F's – one's fantasies, feelings and fun (Holbrook and Hirschman, 1982). They are perceived as abstract and subjective, and mainly refer to an intrinsic motivation in doing something that is inherently interesting and enjoyable (Ridings and Gefen, 2004; Johnson and Ambrose, 2006). Many DECs give users interactive entertainment opportunities and offer them an interesting ambiance. Users of Bebo BeInspired for one example of many, enjoy showing and sharing their personal talents of acting, singing, playing music, etc. Another example is related to entertainment applications on Bebo such as The Simpsons and Pirates Rangers quizzes and games.

*4.1.3 EPISTEMIC VALUE*

Epistemic value can be defined as that value that would persuade users looking for novelty experience as well as new knowledge acquisition (Sheth et al., 1991; Al-Debei and Fitzgerald, 2010). This new knowledge might be derived from different factors of motivations. Many individuals may snoop around in a passive manner for the purpose of sneeking on personal profiles, looking at pictures, and having an eye on what is going on (i.e. interpersonal needs), or may passivley pitch in for the purpose of seeking information, support, advice without intending to publicly engage (i.e. informational needs). Epistemic value is considered a key function of value that is highly related to individuals inner personalities and also can influence behavioral intentions and switch user behaviours (Zeithaml et al., 1996).

*4.1.4 GIFT VALUE*

The huge amounts of random information available on the Internet are staggering. In the world of DECs, the gift value is referred to the public informational products available for everyone at no price with no favor asked in return (Kollack, 1999). In other words, DECs represent a world of tacit information rather than physical objects. DECs are a great source of valuable information with large numbers of users pitching in for the greater good where there is no limit for possibilities. Hence, personal interactions amongst DECs users are best represented as a "gift economy" (Rheingold, 1993). Gift economies are driven by social relations where sharing and exchanging information cost nothing. Nevertheless, the key to a sustainable gift economy lies within the genuine givers who pass on free advice and information to unknown beneficiaries whom they might not even come across again.

*4.1.5 UTILITARIAN VALUE*

Satisfying a utilitarian value is the effective achievement of a functional goal which is often suitable for solution seekers and problem-solvers (Holbrook and Hirschman, 1982). It is characterized as instrumental and extrinsic, that is beneficial for functional and practical queries (Babin et al., 1994). Such values can be classified as Instrumental or Functional which can be best described as an acquisition of new knowledge, and an increase in idea creation and enhanced problem solving (Arguello et al., 2006). For example, when users ask for a handy advice in solving a dilemma related to health, careers, travel, and other issues as in Bebo Young Scot InfoLine, they are seeking for practical, utilitarian values.

Figure 1 below depicts a graphical representation of the above mentioned values as they were thoroughly anayzed within the literature of DECs.

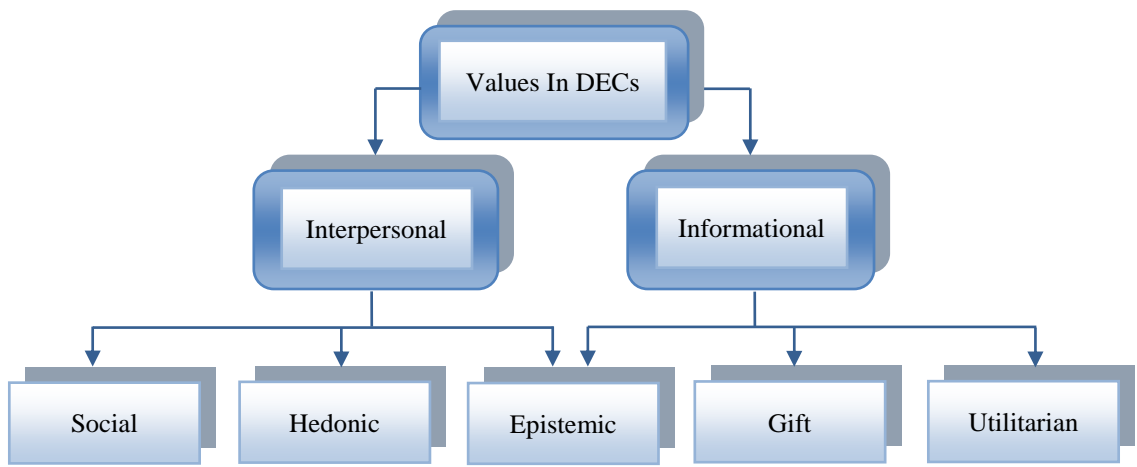

*Figure 1. Values in DECs.*

It is worth mentioning here that such a classification of values in DECs is novel as well as their distinction to *interpersonal values* (social, hedonic and epistemic) and *informational* (epistemic, gift and utilitarian). The same is true concerning the roles adopted by different users of DECs which are unequivocally clarified in the next section.

**4.2 ROLES OF USERS IN DIGITALLY ENGAGED COMMUNITIES**

In time, when enough members join a community, an identity for the community begins to develop. Users might evolve in terms of their participative roles in DECs where each role is distinct with its own characteristics of needs of value. Members start using a common language and as the community grows, they behave according to their intended needs. Then participant roles become identifiable. Some members lead discussions and volunteer information, while others follow, and lurk for support and information. These characteristics, which are common to both online and physical communities, initiate the growth stage of a healthy community (Ridings *et al.,* 2006). Our ethngraphy analysis in this research explores that roles of users in DECs can be broadly classified into:

(1) **Passive users** that are subjected to an action without responding or initiating in return as they flow for self-sake rather than benefiting others. In this research, *Newbie* and *Lurker* are two identifiable roles belong to this calss of users in DECs.

(2) **Active users** that are energetically active in terms of contribution and information sharing. In our context, users in DECs might move from one role to another, or stick to the same role for own self-beneficiary. *Novice*, *Insider*, and *Leader* are three identified roles in DECs that belong to this class of users.

However, recognizable roles within these two broad classes, based on the applied enthnographic analysis, are further discussed below.

*4.2.1 NEWBIE*

A newbie refers to a new comer in any Internet-based activity, most widely used to express newly joining, first-time users of DECs. Newbies start as being observers or over-hearers in order to grab a sense of the community. As they get more familiar with the space, they bring new ideas for discussion and their roles eventually change (Nonnecke and Preece, 2001; Burkett, 2006). They start indirectly by participating through watching or reading information without contributing to the community (Schoberth *et al.*, 2003). But further on, as they have the desire to contribute, they normally become much of contributors.

*4.2.2 LURKER*

Lurkers are depicted as non-contributors, and resource-takers. This is because their main role is observing the community and viewing contents with unstructured levels of participation, and mainly no desire or intention for contributing. They actually do not add any content or engage into any discussion (Nonnecke and Preece, 2000). And if they do, they tend to engage anonymousely. Lurkers are attracted to DECs because of their desire for credible information. They snoop into the community seeking opportunities to broaden their viewpoints and consume information for their own beneficiaries. Approximately and generally speaking, they represent 80–90% of a DEC population (Tedjamulia et al., 2005). Despite the argument of some researchers (e.g. Li et al., 2008) that lurkers are not necessarily passive participants, we agree with Nonnecke and Preece (2000) that lurkers are passive actors as being non-contributors.

*4.2.3 NOVICE*

A novice is a relatively new member of a DEC, who is still inexperienced with patterns of participation. They are beginners who just started to engage within the community. In other words, it is the stage that often follows being a newbie. Once they get fully engaged, they are most likely to contribute on a higher level. Based on that, they are heading towards full participation (Lave and Wenger, 1991). Novice users provide content and tentatively interact in few discussions, post videos, and may comment on others. Novice users as neither lurkers nor leaders and they have been once newbies or lurkers (Kim, 2000).

*4.2.4 INSIDER*

Insiders are regular participants who are fully engaged and committed to the community. They consistently add content and get engaged into group discussions. They can be classified as experienced users as well. Their level of interaction is high and frequent. Insiders make concerted efforts to comment and rate others. They not only browse and ask questions, but respond to others' queries, engage in social interaction, and make intelligent contributions (Tedjamulia et al., 2005). Insiders were once novices (Kim, 2000), but now are established in the community and comfortably participating in the community's ongoing life.

*4.2.5 LEADER*

Leaders can be referred to as key or advanced users. Leaders are defined as contributors to the success and health of the community since they are in a position to spread knowledge, and thus provide cohesiveness and consistency among others. They are the main information providers (Li et al., 2008) as users turn to them for help and thus such users can also be viewed as community moderators. They sustain membership through continuous participation; therefore, they have become recognized within the community. This type of participation is referred to as a "veteran" of a DEC (Tedjamulia et al., 2005) highlighting the fact that they are firmly responsible for making the majority of contributions in the community. The contributions of leaders signifies the main motive fo lurkers to sneak around and decide to get involved (Preece, 2000).

Based on the aforementioned discussion, it is now more clear that users play various roles in DECs, and that behind each role lies a purposive personal desire. However, deciding which role to play might vary depending on how much users are satisfied with meeting their intended values. Similarly to section 4, here in Table 2 we briefly provide further examples of the types of users analyzed in DECs.

Noticeably, users play various roles in DECs, and behind each role lies a purposive personal desire. These roles might vary depending on how much users are satisfied with meeting their intended values.

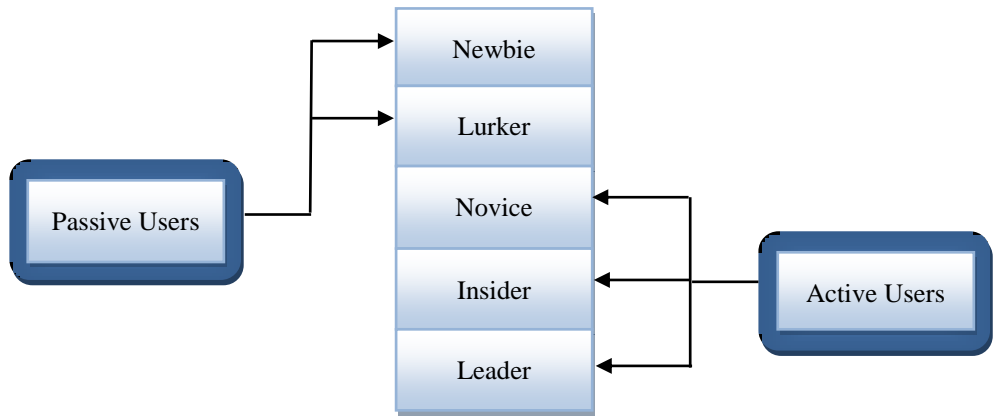

*Figure 2. Roles in DECs.*

Similarly to section 4.1 where the values where summarised, here in figure 2 we have a graphical representation of the various types of users involved in DECs who are mainly divided to passive and active users.

## 5 DISCUSSION AND CONCLUSIONS

The main aim of this paper is to undertsand the different roles of users in DECs and the value elements they capture by playing these roles. To this end, this paper employs an online ethnography on Bebo digitally-engaged community. Based on that, this paper develops a framework that includes two main taxonomies: (1) a taxonomy of values exchanged in DECs; and (2) a taxonomy of roles played by users in DECs in order to statify their needs.

The analysis conducted in this reseach reveals five value elements exchanged in DECs: (a) Social (i.e. emotional, networking, self-esteem, and self-discovery); (b) Hedonic (i.e. self entertainment); (c) Utilitarian Values (i.e. instrumental values); (d) Gift (i.e. free public information); and the (e) Epistemic Values (i.e. acquiring new knowledge). Moreover, this research reveals that users in DECs can be usefully classified as (a) Newbies (i.e. newcomers), (b) Lurkers (i.e. non-contributors), (c) Novices (i.e. beginners), (d) Insiders (i.e. regulars), and (e) Leaders (i.e. experts). Interestingly, these taxonomies seem to be greatly interrelated, as graphically demonstarted in Figure 3. In the researchers' point of view, each role is associated with certain values and vice versa. For example, it comes into view that lurkers might be mainly linked to gift and epistemic value elements, whilst leaders seem to be tightly related to social values in terms of status and self-esteem and so on. The examination of these relationships is however the next step of our research.

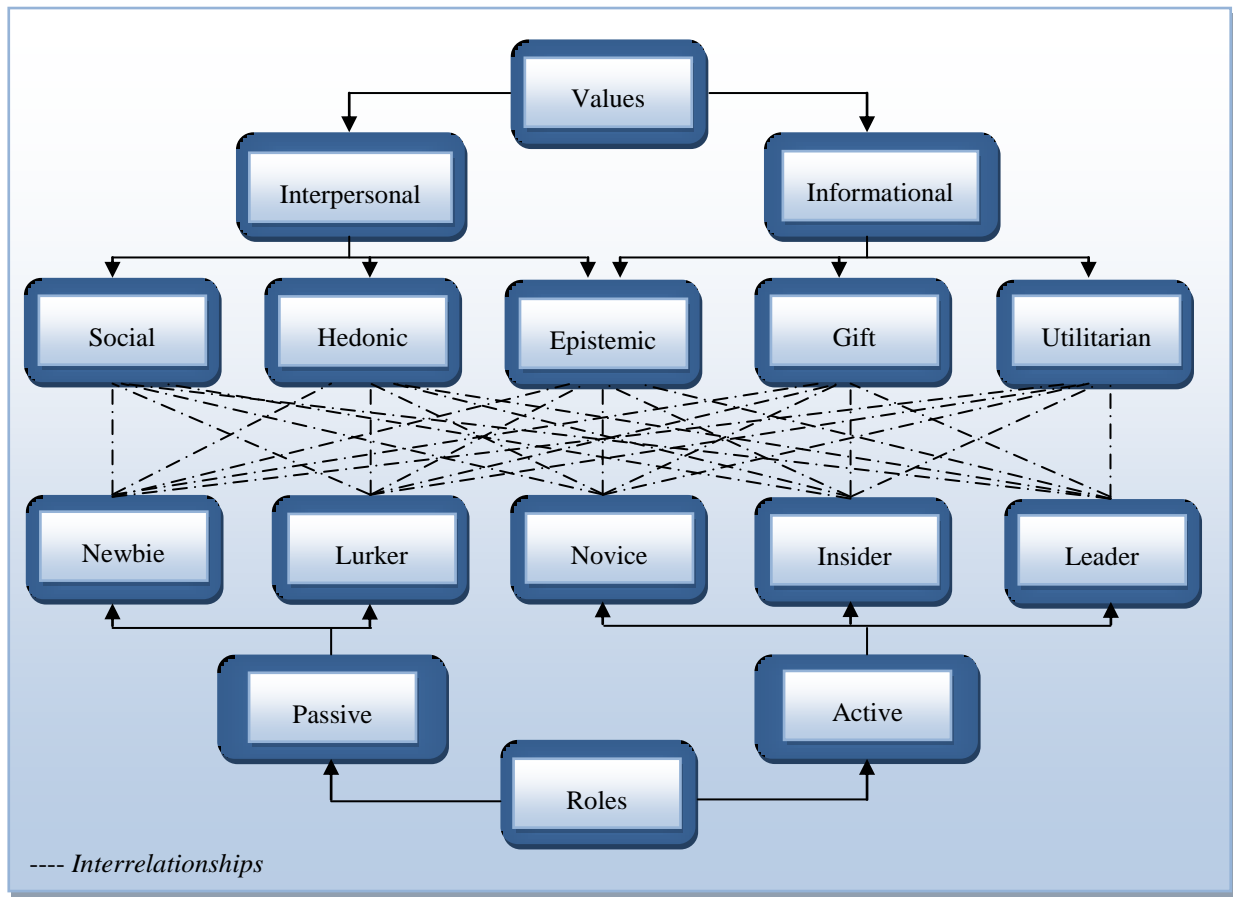

*Figure 3. Inter-relationships between Value Elements and Roles in DECs .*

This paper offers future implications for both theory and practice. From a theoretical perspective, the originality of the framework proposed adds a new dimension of research in DECs, and opens up opportunities for possible extensions and amendments of efforts within this research area. As it develops a comprehensive taxonomy classifying the potential value elements driving users into participating and engaging in DECs as they are expected to be gained and achieved as a result, and develops an inclusive categorization of the various roles adopted and played by users of DECs. From a practical perspective, the study provides insights for:

- *Decision and Policy Makers* in guiding them into identifying their audience and knowing whom to support. Further more, in building strategic plans for a sustainable healthy community, where participation and engagement is continuous, and according to that, policies and regulations might need re-engineering for the sake of supporting certain members.
- *Service Providers* in knowing what factors to examine, whom to support and whom to watch, it eases up their ability of recognizing which parts of the digitally engaged community to balance and to focus on for re-enhancement purposes.
- *Users* where they can exactly know which beenfit and value element they would gain and acquire when acting upon a certain role and vice versa, depending on different situational factors accompanied by every person.
- *Developers*, it inspires them in knowing and meeting the exact needs and intentional values of members according to their different behavioural roles. Taking into consideration the attitudinal, societal, and control differences affecting their intentions and actual behaviours. But as individuals' needs and desires change in each stage of the online community evolution over time, developers require re-designing the tools, features, mechanisms, and technologies. They have to identify carefully each behavioural role played within the community, and know what kind of intentional value elements are related to it, and thus add the right technology components that will better support the community, in a way a sustainable information system life cycle prescribes.